\documentclass[twocolumn,superscriptaddress,amsmath,amssymb,pra,longbibliography]{revtex4-1}
\usepackage{amssymb,amsmath}
\usepackage{cmap}
\usepackage{graphicx}
\usepackage{bm}% bold math
\usepackage{color}
\usepackage{blindtext}
\usepackage{hyperref}
\usepackage{listings}
\usepackage{booktabs}
\usepackage{multirow}
\usepackage{amsmath}
\usepackage{mathrsfs}
\usepackage{braket}

\usepackage{dsfont}
\usepackage{tabularx}

\newcommand{\red}[1]{{\color{black}{#1}}}

\begin{document}

\title{Transverse Shifts and Time Delays of Spatiotemporal Vortex Pulses \\ Reflected and Refracted at a Planar Interface}

\newcommand{\affilB}{Universit\'{e} de Bordeaux, CNRS, LOMA, UMR 5798, Talence, France}
\newcommand{\affilRIKEN}{Theoretical Quantum Physics Laboratory, RIKEN Cluster for Pioneering Research, Wako-shi, Saitama 351-0198, Japan}
\newcommand{\affilM}{CNRS, Centrale Marseille, Institut Fresnel, Aix Marseille University, UMR 7249, 13397 Marseille CEDEX 20, France}

\author{Maxim Mazanov}
\affiliation{V. N. Karazin Kharkiv National University, Kharkiv, 61022, Ukraine}

\author{Danica Sugic}
\affiliation{\affilRIKEN}

\author{Miguel A. Alonso}
\affiliation{\affilM}
\affiliation{The Institute of Optics, University of Rochester, Rochester, NY 14627, USA}

\author{Franco Nori}
\affiliation{RIKEN Center for Quantum Computing, Wako-shi, Saitama 351-0198, Japan}
\affiliation{\affilRIKEN}
\affiliation{Physics Department, University of Michigan, Ann Arbor, Michigan 48109-1040, USA}

\author{Konstantin Y. Bliokh}
\affiliation{\affilRIKEN}

%\date{\today}

\begin{abstract}
Transverse (Hall-effect) and Goos--H\"{a}nchen shifts of light beams reflected/refracted at planar interfaces are important wave phenomena, which can be significantly modified and enhanced by the presence of intrinsic orbital angular momentum (OAM) in the beam. Recently, optical spatiotemporal vortex pulses (STVPs) carrying a purely transverse intrinsic OAM were predicted theoretically and generated experimentally. Here we consider the reflection and refraction of such pulses at a planar isotropic interface. We find theoretically and confirm numerically novel types of the OAM-dependent transverse and longitudinal pulse shifts. Remarkably, the longitudinal shifts can be regarded as time delays, which appear, in contrast to the well-known Wigner time delay, without temporal dispersion of the reflection/refraction coefficients. Such time delays allow one to realize OAM-controlled slow (subluminal) and fast (superluminal) pulse propagation without medium dispersion. These results can have important implications in various problems involving scattering of localized vortex states carrying transverse OAM.  
\end{abstract}

%\keywords{Acoustic force; canonical momentum; acoustic spin; acoustic torque.}

\maketitle

%%%%%%%%%%%%%%%%%%%%%%%%%%%%%%
\section{Introduction}
%{\it Introduction.---}
%%%%%%%%%%%%%%%%%%%%%%%%%%%%%%
Small wavepacket shifts and time delays are currently attracting considerable attention due to their noticeable roles in nanoscience. The first example of such effects is the Goos--H\"{a}nchen shift of the beam reflected/refracted at a planar interface \cite{Goos1947, Artmann1948, Merano2009, Jayaswal2013, Bliokh2013}. This shift is proportional to the wavevector-gradient of the logarithm of the reflection coefficient. 

The temporal counterpart of this spatial shift is the Wigner time delay of a wavepacket scattered by a frequency-dependent potential \cite{Wigner1954,Chiao1997,Carvalho2002,Winful2006,Asano2016}. 
Correspondingly, this delay is given by the frequency gradient of the logarithm of the scattering coefficient. 

Another example of beam shifts is the transverse \red{Imbert--Fedorov shift associated with the spin-Hall effect (i.e., a transverse circular-polarization-induced shift of the reflected/refracted beam}) 
\cite{Imbert1972,Schilling1965,Fedoseyev1988,Onoda2004,Bliokh2006, Hosten2008,Bliokh2013,Gotte2012,Toppel2013,Bliokh2015NP,Bliokh2016,Ling2017}. 
This shift has a more complicated origin associated with the spin angular momentum carried by the wave, spin-orbit interaction, and conservation of the total angular momentum component normal to the interface.

All these shifts and time delays have been studied mostly for Gaussian-like wavepackets and beams, and all have a typical scale of the wavelength or wave period, which can be enhanced up to the beam-width or pulse-length scale using the weak-measurement technique 
\cite{Hosten2008, Gotte2012, Toppel2013, Jayaswal2013, Bliokh2016, Asano2016}. 

It has also been shown that the beam shifts can be modified significantly by the presence of the intrinsic orbital angular momentum (OAM) in optical vortex beams 
\cite{Fedoseyev2001, Dasgupta2006, Fedoseyev2008, Okuda2008, Bliokh2009, Bekshaev2009, Merano2010, Dennis2012, Bliokh2012PRL, Bliokh2013}. 
This enhances the Gaussian-beam shifts by the factor of the OAM quantum number $\ell$ and also produces new types of shifts. 

To the best of our knowledge, the role of the intrinsic OAM and vortices on time delays have not been studied so far. This is because optical vortex beams are usually monochromatic states unbounded in the longitudinal direction, while time delays make sense only for finite-length wavepackets. 

Recently, a novel type of localized pulses carrying transverse intrinsic OAM \red{--- spatiotemporal vortex pulses (STVPs) --- was described theoretically \cite{Sukhorukov2005, Dror2011, Bliokh2012, Bliokh2021} and generated experimentally \cite{Jhajj2016,Hancock2019,Chong2020,Hancock2020,Wan2021,Wang2021} (see also Ref.~\cite{Dallaire2009} for the zeroth-order Bessel STVP without OAM). Such STVPs} have geometrical and OAM properties different from monochromatic vortex beams. \red{(Note that STVPs should not be confused with principally different space-time wavepackets considered in Refs.~\cite{Kondakci2017,Kondakci2019,Turunen2009}.)} Therefore, it is natural to expect that these qualitatively new objects behave differently in problems involving beam shifts and time delays. 

In this work, we consider reflection and refraction of an optical STVP at a planar isotropic interface. We predict theoretically and confirm numerically a number of novel spatial shifts and time delays that are controlled by the value and orientation of the intrinsic OAM of the pulse. Remarkably, time delays appear in this system without any frequency dependence of the reflection/refraction coefficients, thereby allowing one to realize slow (subluminal) and fast (superluminal) pulse propagation without medium dispersion. This is in sharp contrast to Wigner time delays and is produced by the coupling of spatial and temporal degrees of freedom in spatiotemporal vortices. Our results can have important implications in various problems involving scattering of localized vortex states with transverse OAM, both classical and quantum.  
 
%%%%%%%%%%%%%%%%%%%%%%%%%%%%%%
\section{Laguerre-Gaussian STVPs}
\begin{figure}[!t]
\includegraphics[width=\linewidth]{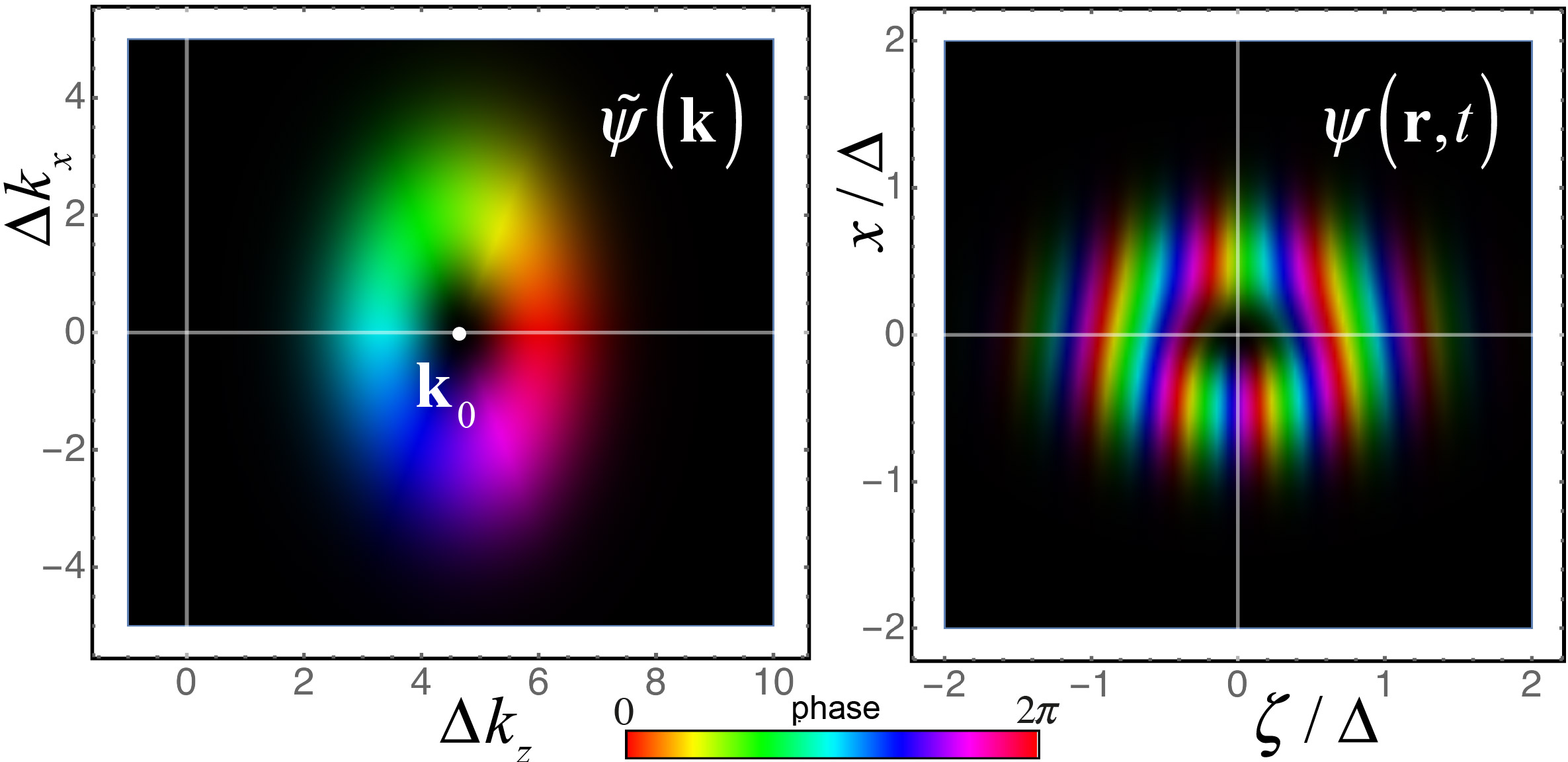}
\caption{
The phase-intensity distributions of the momentum-space (left)
%, $\tilde\psi ({\bf k})$ (a), 
and real-space (right)
%, $\tilde\psi ({\bf r},t)$ (b), 
wavefunctions of the STVP (\ref{eq1}) and (\ref{eq2}) with $\ell=1$, $k_0\Delta = 0.7$ and $\gamma = 1.5$. The brightness is proportional to the intensity $|\psi|^2$, while the color indicates the phase ${\rm Arg}(\psi)$ 
\cite{Thaller_book}.
\label{Fig1}}
\end{figure}

%{\it  Laguerre-Gaussian STVPs.---}
%%%%%%%%%%%%%%%%%%%%%%%%%%%%%%
We first introduce a STVP propagating along the $z$-axis and carrying transverse OAM along the $y$-axis. For this, akin to monochromatic Laguerre-Gaussian (LG) beams \cite{Allen_book,Andrews_book}, we consider a LG-type plane-wave spectrum in the $(z,x)$ plane with the central wavevector ${\bf k}_0 = k_0 \bar{\bf z}$ (the overbar denotes the unit vector of the corresponding axis) and zero radial quantum number (Fig.~\ref{Fig1}):
\begin{equation}
\label{eq1}
\tilde\psi \left( {\bf{k}} \right) \propto {\left[ {\gamma \left( {{k_z} - {k_0}} \right) + i\,{\rm sgn} ( \ell ){k_x}} \right]^{\left| \ell  \right|}}
e^{ - \tfrac{\Delta^2}{4}\left[ {{\gamma ^2}{{\left( {{k_z} - {k_0}} \right)}^2} + k_x^2} \right]} .
\end{equation}
Here, $\ell$ is the integer vortex charge, $\gamma$ is the factor determining the ellipticity of the STVP profile in the $(z,x)$ plane, and $\Delta$ is the $x$-width of the pulse ($\gamma\Delta$ being its $z$-length). Note that we do not include a distribution over $k_y$, because for our goals it is sufficient to consider pulses unbounded along the OAM direction. If needed, an additional Gaussian distribution over $k_y$ can provide localization along the $y$-axis.  

The real-space form of the STVP (\ref{eq1}) is given by the Fourier integral 
$\psi \left( {{\bf{r}},t} \right) \propto \iint {\tilde \psi \left( {\bf{k}} \right){e^{i{\bf{k}} \cdot {\bf{r}} - i\omega \left( {\bf{k}} \right)t}}} d{k_x}d{k_z}$,
where $\omega \left( {\bf{k}} \right) = kc$. 
For the purpose of this work it is sufficient to use a paraxial approximation, $k_0 \Delta \gg 1$, in which only linear deviations in the transverse wavevector components are considered. This leads to the following expression for a paraxial LG-type STVP where diffraction is ignored (Fig.~\ref{Fig1}):
%In the paraxial case $\Delta k_0 \gg 1$ and diffractionless approximation, we obtain an LG-type STVP:
%
\begin{equation}
\label{eq2}
\psi  \propto {\left[ {{\gamma ^{ - 1}}\zeta  + i\,{\rm sgn} ( \ell  )x} \right]^{\left| \ell  \right|}}\!
\exp\! \left[ { - \frac{\left( {{\gamma ^{ - 2}}{\zeta ^2} + {x^2}} \right)}{{{\Delta ^2}}} + i{k_0}\zeta } \right]\! ,
\end{equation}
where $\zeta = z - ct$ is the pulse-accompanying coordinate.
Closed-form real-space expressions that incorporate diffraction both in the paraxial and nonparaxial regimes will be described in a separate work.

%%%%%%%%%%%%%%%%%%%%%%%%%%%%%%%%%%%%%%
%%%%%%%%%%%%%%%%%%%%%%%%%%%%%%%%%%%%%%

%Figure~\ref{Fig1} shows an example of the plane-wave spectrum and real-space profile of the STVP (\ref{eq1}) and (\ref{eq2}). 
For our purposes, the key features of such STVPs are: (i) their spatiotemporal vortex structure near the center: 
$\psi  \propto {\left[ {{\gamma ^{ - 1}}\zeta  + i\,{{\rm sgn}}\! \left( \ell  \right)x} \right]^{\left| \ell  \right|}} e^{ik_0\zeta}$
, and (ii) their normalized integral intrinsic OAM \cite{Bliokh2012, Bliokh2021}: 
\begin{equation}
\label{eq3}
\left\langle {{\bf L}} \right\rangle = \frac{\iint{{\rm Im} \left[ {{\psi ^*}({\bf r} \times\! {\bm \nabla})\psi } \right]}\,dxdz}{\iint{{\psi ^*}\psi}\,dxdz}
%\left\langle {{\bf l}} \right\rangle \equiv
%\frac{{{\omega _0}\left\langle {{\bf L}} \right\rangle }}{{\left\langle I \right\rangle }} 
= \frac{{\gamma  + {\gamma ^{ - 1}}}}{2}\,\ell\, \bar{\bf y} \,.
\end{equation}
% 
%Here, ${\bf L} \propto \omega _0^{ - 1}{\rm Im} \left[ {{\psi ^*}({\bf r} \times {\bm \nabla})\psi } \right]$, $I \propto {\psi ^*}\psi$, 
%$\varphi$ is the azimuthal angle in the $(\zeta,x)$ plane, 
%and $\left\langle {...} \right\rangle  = \iint {...} \,dxdz$.

The above equations are written for a scalar wavefunction $\psi$. To consider polarized optical STVP, one has to multiply each plane wave in the spectrum (\ref{eq1}) by the corresponding electric-field polarization vector ${\bf e}({\bf k}) \bot {\bf k}$. In the paraxial regime this does not affect the shape of the pulse and its OAM considerably, but polarization plays a crucial role in the Goos-H\"{a}nchen and spin-Hall effects \cite{Bliokh2013,Gotte2012,Toppel2013}.

%%%%%%%%%%%%%%%%%%%%%%%%%%%%%%
\section{Reflection/refraction of a STVP at an interface}
\begin{figure*}
\includegraphics[width=0.8\linewidth]{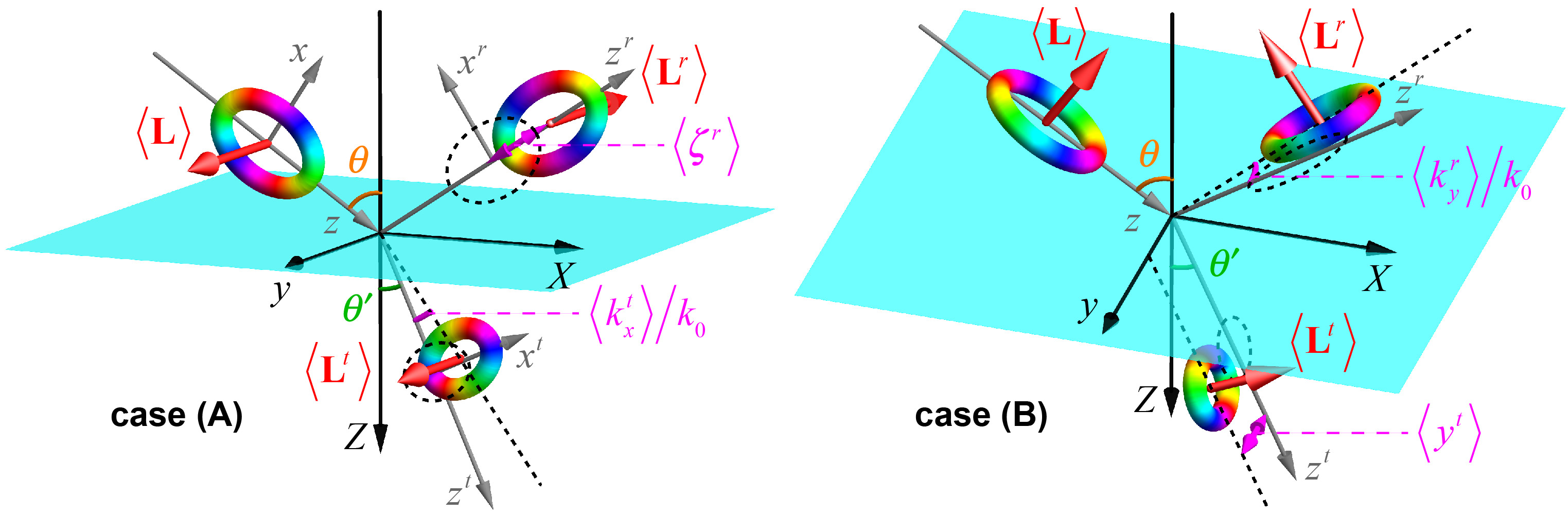}
\caption{
Schematics of the reflection and refraction of a STVP at a planar interface. The incident, reflected, and transmitted pulses, together with their accompanying coordinate frames and intrinsic OAM are shown schematically for the two geometries (A) and (B) (see details in the text).
The longitudinal shift (time delay) $\langle \zeta \rangle$ and angular shift $\langle k_x \rangle$ are shown for the reflected and transmitted pulses in (a), whereas the transverse shift $\langle y \rangle$ and angular shift $\langle k_y \rangle$ are shown for the transmitted and reflected pulses in (b).
%
%The plane-wave spectra (left) and phase-intensity distributions of real-space wavefunctions $\psi({\bf r},t)$ (right) for (a) the monochromatic Bessel beam with $\ell=2$ and (b) spatiotemporal Bessel pulse with $\ell=2$, Eqs.~(\ref{eq1}) and (\ref{eq2}).
%In real-space distributions, the brightness is proportional to the intensity $|\psi|^2$, while the color indicates the phase ${\rm Arg}(\psi)$. 
\label{Fig2}}
\end{figure*}

%{\it  Reflection/refraction of a STVP at an interface.---}
%%%%%%%%%%%%%%%%%%%%%%%%%%%%%%
We now consider reflection/refraction of a paraxial STVP at a planar isotropic (e.g., dielectric) interface. 
%The problem is described the standard Fresnel and Snell formulas applied for each plane wave in the pulse spectrum. Interference between the reflected/refracted plane waves results in small shifts of the resulting pulses as compared to the geometrical-optics expectations. 
The geometry of the problem is shown in Fig.~\ref{Fig2}. The interface is $Z=0$, with the $Z$-axis being directed towards the second medium. The propagation direction of the incident pulse is determined by the central wavevector ${\bf k}_0 = k_0 (\bar{\bf Z}\cos\theta  + \bar{\bf X} \sin\theta) \equiv k_0 \bar{\bf z}$. According to Snell's law, the reflected and transmitted pulses have the central wavevectors ${\bf k}^r_0 = k_0 (- \bar{\bf Z}\cos\theta  + \bar{\bf X} \sin\theta)\equiv k_0 \bar{\bf z}^r$ and ${\bf k}^t_0 = k_0^\prime ( \bar{\bf Z}\cos\theta^\prime  + \bar{\bf X} \sin\theta^\prime) \equiv k_0^\prime \bar{\bf z}^t$ ($\sin\theta^\prime = n^{-1} \sin\theta$, $k_0^\prime = n k_0$, where $n$ is the relative refractive index of the second medium), respectively. 
Here, as usual in beam-shift problems, we use the accompanying coordinate frames $(x,y,z)$ and $(x^{r,t},y,z^{r,t})$ for the incident and reflected/transmitted pulses, Fig.~\ref{Fig2}. 

%%%%%%%%%%%%%%%%%%%%%%%%%%%%%%%%%%%%%%
%%%%%%%%%%%%%%%%%%%%%%%%%%%%%%%%%%%%%%

In contrast to the monochromatic-beam-shift problems, where the orientation of the OAM is fixed by the beam propagation direction, in our problem the transverse OAM can have different orientations with respect to the $(x,z)$ plane of incidence. We will consider two basic cases shown in Fig.~\ref{Fig2}: 

(A) The incident STVP is localized in the $(x,z)$ plane, and the intrinsic OAM $\left\langle {{\bf L}} \right\rangle \parallel \bar{\bf y}$.  

(B) The incident STVP is localized in the $(y,z)$ plane and $\left\langle {{\bf L}} \right\rangle \parallel \bar{\bf x}$.  

To describe the main transformations of the reflected and refracted STVPs, note that the $y$-components of the wavevectors in their spectra are conserved, $k_y^{r,t} = k_y$, while the $x$-components in the corresponding accompanying frames are related as $k_x^r = - k_x$ and $k_x^t = (\cos\theta/\cos\theta^\prime) k_x$ \cite{Bliokh2013}. In addition, the $z$-components of the wavevectors of the transmitted pulse are $k_z^t \simeq n k_z$. From this, one can find that the vortex is inverted in the reflected pulse in the case (A) but not (B), and its intrinsic OAM becomes (see Fig.~\ref{Fig2}):  
\begin{align}
\label{eq4}
\left\langle {{{\bf{L}}^r}} \right\rangle_{A}  =  - \left\langle {\bf{L}} \right\rangle  =  - \frac{{\gamma  + {\gamma ^{ - 1}}}}{2}\,\ell \, {\bf{\bar y}} \,,
%~~{\rm case ~(A)}
\nonumber \\
\left\langle {{{\bf{L}}^r}} \right\rangle_{B}  =   \frac{{\gamma  + {\gamma ^{ - 1}}}}{2}\,\ell \, {\bf{\bar x}}^r .~~~
% ~~{\rm case ~(B)}
\end{align}
Here and hereafter, the subscripts $A$ and $B$ mark the quantities related to the cases (A) and (B), respectively.

For the transmitted STVP, the above transformations of the wavevector components stretch the $x^t$-width of the pulse by a factor of $\cos\theta^\prime/\cos\theta$ and squeeze its longitudinal length by a factor of $1/n$. Therefore, the intrinsic OAM of the refracted pulses becomes
\begin{align}
\label{eq5}
\left\langle {{{\bf{L}}^t}} \right\rangle_{A}  =  \frac{{\gamma'_A  + {\gamma'^{- 1}_A}}}{2}\,\ell \, {\bf{\bar y}} ,~~
\gamma'_A = \frac{\cos\theta}{n\cos\theta^\prime}\gamma,
%~~{\rm case ~(A)}
\nonumber \\
\left\langle {{{\bf{L}}^t}} \right\rangle_{B}  =   \frac{{\gamma'_B  + {\gamma'^{- 1}_B}}}{2}\,\ell \, {\bf{\bar x}}^t ,~~~
\gamma'_B = \frac{\gamma}{n}.~~~
%~~~{\rm case ~(B)}
\end{align}
% 
%where $\gamma^\prime = n^{-1}(\cos\theta/\cos\theta^\prime)\gamma$ and $\gamma'' = n^{-1}\gamma$ are the ellipticity factors of the refracted pulse in the cases (A) and (B).
\red{Equations (\ref{eq4}) and (\ref{eq5}) show that the transformations of the transverse intrinsic OAM in the case (A) is similar to those of the longitudinal OAM of monochromatic vortex beams \cite{Bliokh2009,Bliokh2013}, only with the additional $n^{-1}$ factor in $\gamma'_A$. In turn, the case (B) differs considerably because the intrinsic OAM and vortex do not flip in the reflected pulse.}

%%%%%%%%%%%%%%%%%%%%%%%%%%%%%%
\section{Transverse shifts and time delays}
%{\it  Transverse shifts and time delays.---}
%%%%%%%%%%%%%%%%%%%%%%%%%%%%%%
We are now in the position to calculate small shifts in reflected/refracted STVPs. Rigorous calculations can be performed by applying   the standard Fresnel-Snell formulas to each plane wave in the incident pulse spectrum; this is realized numerically in the next section. Here, akin to Ref.~\cite{Bliokh2009}, we derive all the OAM-dependent shifts using general considerations. 

First of all, we assume that paraxial polarized optical STVPs experience all the polarization-dependent shifts known for Gaussian wave beams or packets, i.e., angular and spatial Goos-H\"{a}nchen and spin-Hall shifts
 $\left\langle k^{t,r}_x \right\rangle_0$, $\left\langle x^{t,r} \right\rangle_0$,  $\left\langle k^{t,r}_y \right\rangle_0$, and  $\left\langle y^{t,r} \right\rangle_0$ \cite{Bliokh2013,Gotte2012,Toppel2013}, where the subscript ``0'' indicates that the shifts are calculated for Gaussian states with $\ell=0$. 
In addition to these shifts, we will determine the $\ell$-dependent shifts induced by the transverse intrinsic OAM. There are three types of such shifts. 

The first type is related to the conservation of the $Z$-component of the total angular momentum in the problem and can be associated with the {\it orbital-Hall effect of light} \cite{Bliokh2013,Bliokh2006_II}. In the case (A) the intrinsic OAM has only the $y$-component, and the conservation law is satisfied trivially. In the case (B), the incident and reflected pulses have the same $Z$-components of the normalized intrinsic OAM, $\left\langle L_Z \right\rangle = \left\langle L^r_Z \right\rangle$, Eqs.~(\ref{eq3}) and (\ref{eq4}), while the transmitted pulse has a different OAM component: $\left\langle L_Z \right\rangle \neq \left\langle L^t_Z \right\rangle$, Eqs.~(\ref{eq3}) and (\ref{eq5}). Similarly to the refraction of monochromatic vortex beams \cite{Bliokh2013,Bliokh2009,Fedoseyev2008,Merano2010}, this imbalance between the intrinsic OAM of the incident and transmitted pulses should be compensated by the transverse $y$-shift of the refracted pulse producing an extrinsic OAM $\langle L_Z^t \rangle^{\rm ext} = \langle y^t \rangle \langle k_X^t \rangle \simeq \langle y^t \rangle\, nk_0 \sin\theta^\prime$. From here, the refracted STVP in the case (B) should undergo an additional transverse shift (see Fig.~\ref{Fig2})
\begin{equation}
\label{eq6}
\left\langle {{y^t}} \right\rangle_{B}  = \frac{{\left\langle {L_Z^t} \right\rangle  - \left\langle {{L_Z}} \right\rangle }}{{n{k_0}\sin \theta '}} = \frac{{\gamma \ell }}{{2{k_0}}}\left( {{n^{ - 2}} - 1} \right) .
\end{equation}
In contrast to the analogous shift for refracted monochromatic vortex beams, the shift (\ref{eq6}) is independent of the angle $\theta$ (apart from the small vicinity of the normal incidence $\theta =0$, which is singular for the transverse-shift problem). The typical scale of this shift is the wavelength, although it can be enhanced by high vortex charges $\ell$ or ellipticity $\gamma \gg 1$ (narrow long pulses).

The second type of $\ell$-dependent shift is related to the angular Goos-H\"{a}nchen and spin-Hall shifts $\langle k_{x,y} \rangle$, see Fig.~\ref{Fig2}. As has been shown for monochromatic vortex beams, in the presence of vortex these shifts acquire an additional factor of $\left(1+|\ell|\right)$ \cite{Bliokh2009,Bliokh2013,Merano2010}, so that the additional shifts are:
\begin{equation}
\label{eq7}
\left\langle k^{t,r}_x \right\rangle_{A} = |\ell| \left\langle k^{t,r}_x \right\rangle_0, \quad
\left\langle k^{t,r}_y \right\rangle_{B} = |\ell| \left\langle k^{t,r}_y \right\rangle_0\,.
\end{equation}
% 
%These equations are valid for the both cases (A) and (B).
The typical scale of these angular shifts is the inverse Rayleigh range $\sim 1/(k_0 \Delta^2)$, and these shifts are independent of the ellipticity $\gamma$.

Finally, the third type of $\ell$-dependent shifts is related to the cross-coupling between different Cartesian degrees of freedom in a vortex. Below we use reasoning similar to that for vortex beams in Refs.~\cite{Bliokh2013,Bliokh2009}. In the case (A), the spatiotemporal vortices in the reflected and transmitted pulses have the forms $\propto {\left[ { - {\gamma ^{ - 1}}{\zeta ^r} + i\,{\rm sgn}\! \left( \ell  \right){x^r}} \right]^{\left| \ell  \right|}}$ and $\propto {\left[ {{\gamma'^{ - 1}_A}{\zeta ^t} + i\,{\rm sgn}\! \left( \ell  \right){x^t}} \right]^{\left| \ell  \right|}}$, respectively, where $\zeta^{r,t} = z^{r,t} - ct$ and $c$ is the speed of light in the corresponding medium. Among other polarization-dependent shifts, these pulses experience shifts in momentum space due to the angular Goos-H\"{a}nchen effect, which can be regarded as {\it imaginary} shifts in real space \cite{Aiello2008,Bliokh2009,Bliokh2013}: ${\left\langle {k_x^r} \right\rangle _0} \to \delta {x^r} =  - i\dfrac{\Delta^2}{2}{\left\langle {k_x^r} \right\rangle _0}$ and ${\left\langle {k_x^t} \right\rangle _0} \to \delta {x^t} =  - i\dfrac{\Delta^2}{2}{\left( {\dfrac{{\cos \theta '}}{{\cos \theta }}} \right)^2}{\left\langle {k_x^t} \right\rangle _0}$. Substituting these shifts to the vortex forms of the reflected and transmitted pulses, we find that the imaginary $x$-shifts produce real $\ell$-dependent $\zeta$-shifts as follows (see Fig.~\ref{Fig2}):
\begin{equation}
\label{eq8}
\left\langle {{\zeta ^r}} \right\rangle_A  = - \ell\,\gamma \frac{{{\Delta ^2}}}{2}{\left\langle {k_x^r} \right\rangle _0}, ~~
\left\langle {{\zeta ^t}} \right\rangle_A  =  \ell\, \frac{\gamma}{n} \frac{{{\Delta ^2}}}{2}\frac{{\cos \theta '}}{{\cos \theta }}{\left\langle {k_x^t} \right\rangle _0}.
\end{equation}
Applying analogous considerations to the case (B), with the reflected and transmitted vortices $\propto {\left[ {y+ i{\gamma ^{ - 1}}{\rm sgn}\! \left( \ell  \right){\zeta ^r}} \right]^{\left| \ell  \right|}}$ and 
$\propto {\left[ {y+ i{\gamma_B^{\prime - 1}}{\rm sgn}\! \left( \ell  \right){\zeta^t}} \right]^{\left| \ell  \right|}}$, and angular Hall-effect shifts ${\left\langle {k_y^{r,t}} \right\rangle _0} \to \delta y^{r,t} =  - i\dfrac{{{\Delta ^2}}}{2}{\left\langle {k_y^{r,t}} \right\rangle _0}$, where $\Delta$ is the pulse width in the $y$-direction, we obtain 
\begin{equation}
\label{eq9}
\left\langle {{\zeta ^r}} \right\rangle_B  = - \ell\,\gamma \frac{{{\Delta ^2}}}{2}{\left\langle {k_y^r} \right\rangle _0}, ~~
\left\langle {{\zeta ^t}} \right\rangle_B  =  \ell\, \frac{\gamma}{n} \frac{{{\Delta ^2}}}{2}{\left\langle {k_y^t} \right\rangle _0}.
\end{equation}

Equations (\ref{eq8}) and (\ref{eq9}) describe a remarkable qualitatively novel phenomenon: longitudinal shifts of STVPs reflected/refracted by a planar interface. These $\zeta$-shifts are equivalent to {\it time delays} $\left\langle \delta t \right\rangle = - \left\langle {{\zeta}} \right\rangle/c$.
%, where $c$ is the speed of light in the corresponding medium. 
In contrast to the Wigner time delays, produced by the temporal dispersion (frequency  dependence) of the scattered potential \cite{Wigner1954,Chiao1997,Carvalho2002,Winful2006,Asano2016}, 
here the time delays appear without any temporal dispersion. The angular Goos-H\"{a}nchen effect ${\left\langle {k_x^{r,t}} \right\rangle _0}$ originates from the spatial dispersion (wavevector dependence) of the Fresnel reflection/transmission coefficients, while the angular spin-Hall shift ${\left\langle {k_y^{r,t}} \right\rangle _0}$ is a purely geometric phenomenon which does not require any dispersion \cite{Bliokh2015NP}. 

Importantly, such time delays allow one to realize {\it slow} (subluminal, $\left\langle {{\zeta}} \right\rangle <0$) and {\it fast} (superluminal, $\left\langle {{\zeta}} \right\rangle >0$) pulse propagation without any dispersion in optical media. Unlike previous approaches controlling slow/fast light via properties of the medium, we can control propagation time via internal spatiotemporal properties of the pulse. 
\red{Note, however, that, in contrast to the wave packets in Ref.~\cite{Kondakci2019}, the sub- or superluminal propagation of the pulses studied here is induced by the Fresnel-Snell reflection/refraction at an interface rather than by tailoring the pulse to have a free-space group velocity differing from $c$.}

Equations (\ref{eq8}) and (\ref{eq9}) show that these novel OAM-dependent time delays are a rather universal phenomenon: they appear in both reflected and transmitted pulses in both cases (A) and (B). 
It is natural to expect that such time delays will appear in a variety of systems, both classical and quantum, involving scattering of a spatiotemporal vortex with the transverse OAM.

The typical magnitude of the longitudinal shifts (\ref{eq8}) and (\ref{eq9}) is the wavelength. However, angular shifts $\left\langle k^{r}_{x,y} \right\rangle_0$ of the reflected pulses, and hence the corresponding new shifts (\ref{eq7})--(\ref{eq9}), are enhanced resonantly for near-$p$ polarization in the vicinity of the {\it Brewster angle} of incidence $\theta_B = \tan^{-1} (n)$ \cite{Merano2009,Bliokh2006,Dasgupta2006,Qin2009} (see Fig.~\ref{Fig4} below). This is a general phenomenon of the {\it weak-measurement} amplification of shifts for wavepackets scattered with a near-zero amplitude \cite{Asano2016,Gorodetski2012,Gotte2013}.
\red{The maximum weak-measurement-amplified shift is comparable with the pulse size in the corresponding dimension, which corresponds to the amplification factor $
\sim k_0 \Delta  \gg 1$.}

%The problem is described the standard Fresnel and Snell formulas applied for each plane wave in the pulse spectrum. Interference between the reflected/refracted plane waves results in small shifts of the resulting pulses as compared to the geometrical-optics expectations. 

%%%%%%%%%%%%%%%%%%%%%%%%%%%%%%%%%%%%%%
\begin{figure}[!t]
\includegraphics[width=\linewidth]{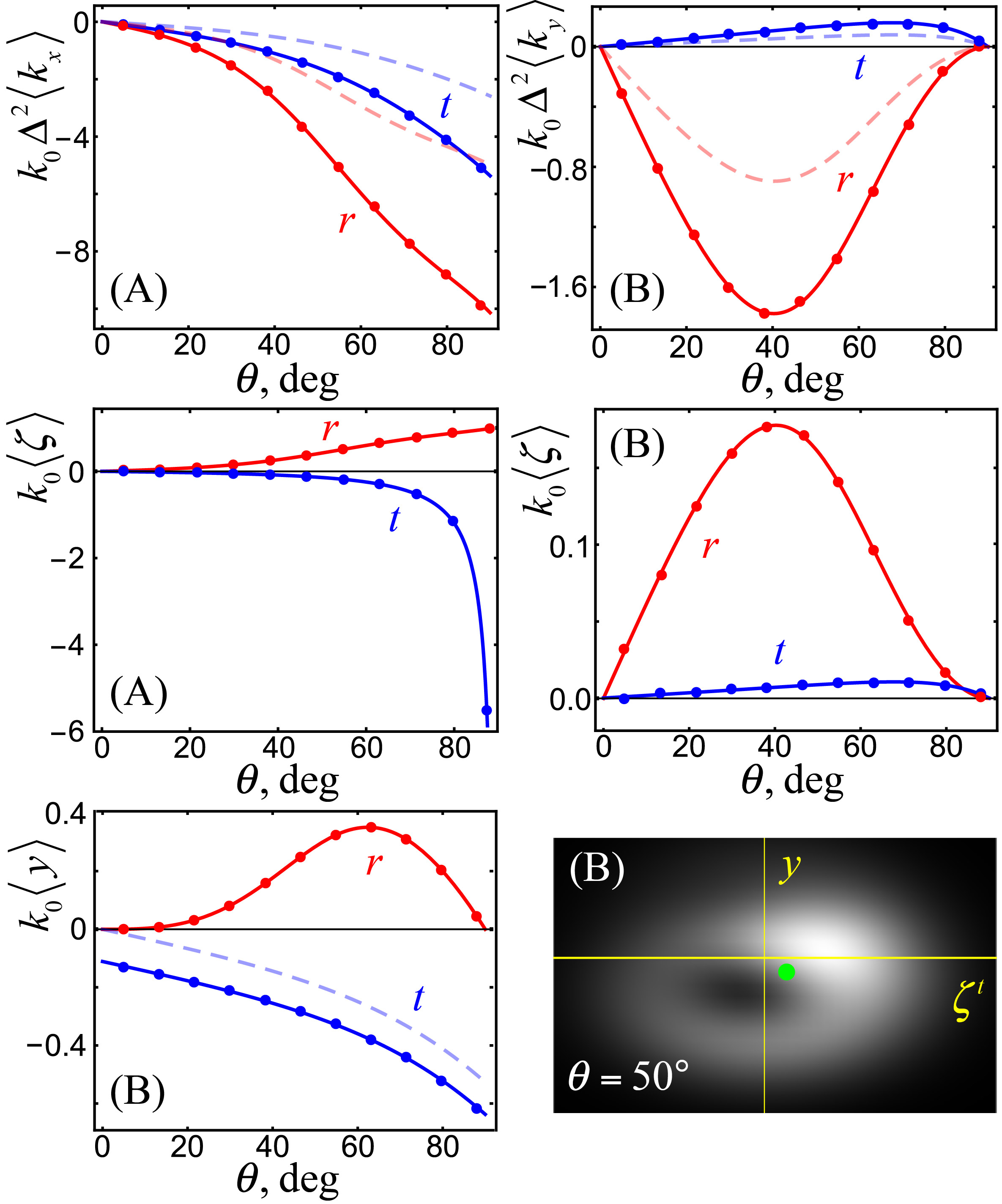}
\caption{
Theoretically calculated (curves) and numerically calculated (symbols) shifts of the reflected ($r$) and transmitted ($t$) STVPs in the cases (A) and (B) from Fig.~\ref{Fig2} as functions of the angle of incidence $\theta$. The shifts are given by the sums of previously known polarization-induced contributions at $\ell = 0$ (shown by pale dashed curves) and OAM-induced contributions Eqs.~(\ref{eq6})--(\ref{eq9}). Parameters are: $\ell = 1$, $n=1.5$, $\gamma = 0.4$, $k_0\Delta = 500$, and ${\bf e}_0 \equiv (e_x,e_y) = \left(1/\sqrt{3}, (1-i)/\sqrt{3}\right)$.
The density plot shows an example of the deformation of the transmitted pulse in the case (B) with $k_0\Delta = 1$ and $\gamma = 2.5$ (to enhance the shifts with respect to the pulse size) and its centroid right after the refraction.
\label{Fig3}}
\end{figure}
%%%%%%%%%%%%%%%%%%%%%%%%%%%%%%%%%%%%%%

%%%%%%%%%%%%%%%%%%%%%%%%%%%%%%
\section{Numerical calculations}
%{\it  Numerical calculations.---}
%%%%%%%%%%%%%%%%%%%%%%%%%%%%%%
To verify the above theoretical derivations, we performed numerical calculations of the reflection/refraction of polarized STVPs at a dielectric interface by applying exact Fresnel-Snell's formulas to each plane wave in the incident pulse spectrum 
$\tilde{\bf{E}}( {\bf{k}} ) = {\bf{e}}( {\bf{k}} )\tilde \psi ( {\bf{k}} )$. In the paraxial approximation, this is equivalent to applying an effective wavevector-dependent Jones matrix ${{\hat T}^{r,t}}( {\bf{k}})$ to the polarization of the central plane wave ${\bf e}_0 = {\bf e} ({\bf k}_0)$ \cite{Bliokh2013,Gotte2012,Toppel2013}, so that the reflected and transmitted pulse spectra become $\tilde{{{\bf{E}}}}^{r,t}( {\bf{k}} ) \simeq {{\hat T}^{r,t}}( {\bf{k}} )\,{\bf{e}}_0\,\tilde \psi ( {\bf{k}} )$. After that, the spatial and angular shifts are calculated as expectation values of the corresponding position and momentum operators in the momentum representation: 
$\left\langle {{y^{r,t}}} \right\rangle  = {{\left\langle {{{\tilde{\bf{ E}}}^{r,t}}} \right|i\partial /\partial k_y^{r,t}\left| {{{\tilde{\bf{ E}}}^{r,t}}} \right\rangle } \mathord{\left/
 {\vphantom {{\left\langle {{{\tilde{\bf{ E}}}^{r,t}}} \right|i\partial /\partial k_y^{r,t}\left| {{{\tilde{\bf{ E}}}^{r,t}}} \right\rangle } {\left\langle {{{\tilde{\bf{ E}}}^{r,t}}} \right.\left| {{{\tilde{\bf{ E}}}^{r,t}}} \right\rangle }}} \right.
 \kern-\nulldelimiterspace} {\left\langle {{{\tilde{\bf{ E}}}^{r,t}}}\! \right.\left| {{{\tilde{\bf{ E}}}^{r,t}}} \right\rangle }}$,
$\left\langle {{\zeta ^{r,t}}} \right\rangle  = {{\left\langle {{{\tilde{\bf{ E}}}^{r,t}}} \right|i\partial /\partial k_z^{r,t}\left| {{{\tilde{\bf{ E}}}^{r,t}}} \right\rangle } \mathord{\left/
 {\vphantom {{\left\langle {{{{\bf{\tilde E}}}^{r,t}}} \right|i\partial /\partial k_z^{r,t}\left| {{{{\bf{\tilde E}}}^{r,t}}} \right\rangle } {\left\langle {{{\tilde{\bf{ E}}}^{r,t}}} \right.\left| {{{\tilde{\bf{ E}}}^{r,t}}} \right\rangle }}} \right.
 \kern-\nulldelimiterspace} {\left\langle {{{\tilde{\bf{ E}}}^{r,t}}}\! \right.\left| {{{\tilde{\bf{ E}}}^{r,t}}} \right\rangle }}$,
$\left\langle {k_{x,y}^{r,t}} \right\rangle  = {{\left\langle {{{\tilde{\bf{ E}}}^{r,t}}} \right|k_{x,y}^{r,t}\left| {{{\tilde{\bf{ E}}}^{r,t}}} \right\rangle } \mathord{\left/
 {\vphantom {{\left\langle {{{{\bf{\tilde E}}}^{r,t}}} \right|k_{x,y}^{r,t}\left| {{{{\bf{\tilde E}}}^{r,t}}} \right\rangle } {\left\langle {{{{\bf{\tilde E}}}^{r,t}}} \right.\left| {{{{\bf{\tilde E}}}^{r,t}}} \right\rangle }}} \right.
 \kern-\nulldelimiterspace} {\left\langle {{{\tilde{\bf{ E}}}^{r,t}}} \right.\left| {{{\tilde{\bf{ E}}}^{r,t}}} \right\rangle }}$,
where the inner product involves integration over the corresponding wavevector components: $\left(k_x^{r,t},k_z^{r,t}\right)$ and $\left(k_y,k_z^{r,t}\right)$ in the cases (A) and (B), respectively. 

In doing so, it is sufficient to use the approximation linear in the transverse wavevector components for all shifts apart from $\left\langle {{y^t}} \right\rangle$, Eq.~(\ref{eq6}). For this shift it is necessary to consider the second-order correction from the Snell's transformation of the wavevectors, see Eqs.~(15) and (19) in Ref.~\cite{Fedoseyev2001} for a similar peculiarity in monochromatic vortex beams. In our case, this second-order correction is given by 
${k_z} - {k_0} \simeq k_z^t - n{k_0} - \dfrac{{k_y^2}}{{2{k_0}}}\left( {{n^{ - 2}} - 1} \right)$. 

\begin{figure}[!t]
\includegraphics[width=\linewidth]{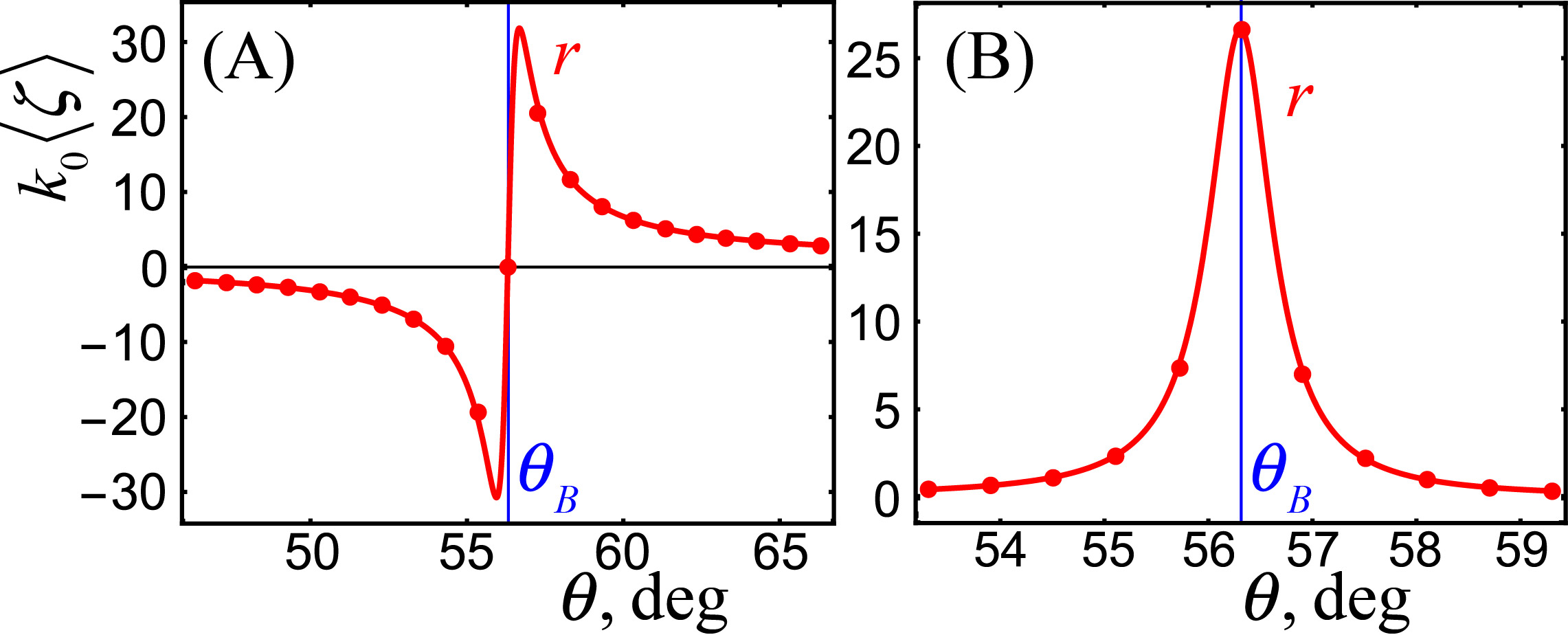}
\caption{
Resonant weak-measurement enhancement of the longitudinal shifts (time delays) of the reflected STVPs for near-$p$ input polarization ${\bf e}_0 = \left(1,0.01\right)$ in the vicinity of the Brewster angle of incidence in the cases (A) and (B). Parameters are: $\ell = 1$, $n=1.5$, $\gamma = 0.4$, $k_0\Delta = 500$.
\label{Fig4}}
\end{figure}

Figures~\ref{Fig3} and \ref{Fig4} display results of numerical calculations of the shifts (\ref{eq6})--(\ref{eq9}) for an air-glass interface with $n=1.5$, generic incident STVP and different angles of incidence $\theta$. These calculations demonstrate perfect agreement with the theoretical predictions.
Furthermore, Fig.~\ref{Fig3} also shows a typical real-space intensity profile of the transmitted STVP which exhibits deformations characteristic for shifts of vortex beams \cite{Bliokh2013,Dennis2012}. Figure~\ref{Fig4} demonstrates weak-meaurement enhancement \cite{Asano2016,Gorodetski2012,Gotte2013}, by two orders of magnitude, of the longitudinal shifts (time delays) $\langle \zeta^r \rangle$ of reflected pulses for a near-$p$ input polarization in the vicinity of the Brester angle of incidence, $\theta \simeq \theta_B$.

%%%%%%%%%%%%%%%%%%%%%%%%%%%%%%%%%%%%%%
%%%%%%%%%%%%%%%%%%%%%%%%%%%%%%%%%%%%%%

%%%%%%%%%%%%%%%%%%%%%%%%%%%%%%
\section{Discussion}
%{\it Discussion.---}
%%%%%%%%%%%%%%%%%%%%%%%%%%%%%%
We have described reflection and refraction of a STVP at a planar isotropic interface. The problem was considered by adopting previously developed methods for monochromatic vortex beams. In doing so, spatiotemporal vortices have a more complicated geometry with a transverse intrinsic OAM, which requires consideration of two basic cases: (A) the OAM is orthogonal to the plane of incidence and (B) the OAM lies within this plane. We have described transformations of the reflected and transmitted pulses in both of these cases. Notably, reflection in the case (A) can be used to flip the intrinsic OAM of the pulse, while refraction can be employed for changing the ellipticity of the pulse.   

Most importantly, we have derived analytically and checked numerically all OAM-dependent spatial and angular shifts of the reflected and transmitted pulses in the paraxial approximation. These shifts can be divided into three types: (i) the spatial orbital-Hall-effect shift $\langle y \rangle$ appearing for the transmitted pulse in the case (B); (ii) the OAM-amplified angular Goos-H\"{a}nchen and Hall-effect shifts $\langle k_x \rangle$ and $\langle k_y \rangle$; and (iii) the longitudinal shifts $\langle \zeta \rangle$ which appear for both reflected and transmitted pulses in both cases (A) and (B). The latter one is the most remarkable phenomenon, which is equivalent to {time delays} $\left\langle \delta t \right\rangle = - \left\langle {{\zeta}} \right\rangle/c$ of the scattered pulses. In contrast to the well-known Wigner time delay, this effect occurs without any temporal dispersion of the scattering coefficients, from the coupling of spatial and temporal degrees of freedom in spatiotemporal vortices.  
Such time delays allow one to realize OAM-controlled slow (subluminal) and fast (superluminal) pulse propagation without any medium dispersion.

Due to remarkable success in experimental studies of subwavelength shifts of monochromatic optical beams and Wigner time delays of optical pulses, it is natural to expect that the new shifts and time delays predicted in this work could be measured in the near future. 
%Importantly, similar to the previously studied monochromatic-beam shifts, the new angular shifts and time delays of STVPs could be dramatically enhanced for reflected pulses with near-$p$ polarization in the vicinity of the Brewster angle of incidence \cite{Merano2009,Bliokh2006,Dasgupta2006,Qin2009}. This is a general phenomenon of weak-measurement amplification of shifts for wavepackets scattered with a near-zero amplitude \cite{Asano2016,Gorodetski2012,Gotte2013}.
Furthermore, our work can stimulate a number of implications and follow-up studies. In particular, scattering of quantum spatiotemporal vortices in the geometry (A) can appear in 2D condensed-matter systems. Furthermore, we have considered only the basic case of STVP with a purely transverse intrinsic OAM and two basic geometries (A) and (B) with respect to the interface. In general, one can examine STVPs with intrinsic OAM arbitrarily oriented with repect to the propagation direction \cite{Bliokh2012,Wan2021} and interface. One can expect that in this general case, the pulse shifts could be expressed via suitably weighted sums of previously considered basic shifts. Finally, including temporal dispersion of the media and interface into consideration should add the Wigner time-delay effects, which could be coupled with spatial degrees of freedom and produce new spatial pulse shifts.  

\vspace*{0.3cm}
\begin{acknowledgments}
%%%%%%%%%%%%%%%%%%%%%%%%%%%%%%
%\section*{ACKNOWLEDGMENTS}
%%%%%%%%%%%%%%%%%%%%%%%%%%%%%%
We are grateful to V. G. Fedoseyev, A. Y. Bekshaev, and O. Yermakov for helpful discussions. 
This work was partially supported by the National Research Foundation of Ukraine (Project No. 2020.02/0149 “Quantum phenomena in the interaction of electromagnetic waves with solid-state nanostructures”) and the Excellence Initiative of Aix Marseille University—A*MIDEX, a French ‘Investissements d’Avenir’ programme. F.N. was supported by Nippon Telegraph and Telephone Corporation (NTT) Research; the Japan Science and Technology Agency (JST) via the Quantum Leap Flagship Program (Q-LEAP), the Moonshot R\&D Grant No. JP- MJMS2061, and the Centers of Research Excellence in Science and Technology (CREST) Grant No. JPMJCR1676; the Japan Society for the Promotion of Science (JSPS) via the Grants-in-Aid for Scientific Research (KAKENHI) Grant No. JP20H00134, and the JSPS–RFBR Grant No. JPJSBP120194828; the Army Research
Office (ARO) (Grant No. W911NF-18-1-0358), the Asian Office of Aerospace Research and Development (AOARD) (Grant No. FA2386-20-1-4069); and the Foundational Questions Institute Fund (FQXi) (Grant No. FQXi-IAF19-06).
\end{acknowledgments}

%\newpage

%\pagebreak

\bibliography{References}

\end{document}